\documentclass[prb,twocolumn,showpacs,preprintnumbers,amsmath,amssymb]{revtex4}

\usepackage{graphicx}
\usepackage{dcolumn}
\usepackage{epsfig}

\begin{document}
\title{Broadband electrical detection of spin excitations in Ga$_{0.98}$Mn$_{0.02}$As using a photovoltage technique}

\author{Andr\'{e} Wirthmann$^{1}$\footnote{Electronic address: wirthmann@physics.umanitoba.ca}, Xiong Hui$^{1,2}$, N. Mecking$^{1},$\footnote{Current address: Institut f\"ur Angewandte Physik und Zentrum f\"ur Mikrostrukturforschung, Universit\"at Hamburg, Jungiusstra\ss e 11, 20355 Hamburg, Germany}, Y. S. Gui$^{1}$, T. Chakraborty$^{1}$, and C.-M. Hu$^{1}$}
\affiliation{$^{1}$Department of Physics and Astronomy, University of Manitoba, Winnipeg, Manitoba R3T 2N2, Canada}
\affiliation{$^{2}$Surface Physics Laboratory, Department of Physics, Fudan University, Shanghai 200433, People's Republic of China}

\author{M. Reinwald, C. Sch\"{u}ller, and W. Wegscheider}
\affiliation{Institut f\"ur Experimentelle und Angewandte Physik, Universit\"at Regensburg, Universit\"atsstra\ss e 31, 93040 Regensburg, Germany}

\begin{abstract}
We report on microwave photovoltage and simultaneous magnetotransport measurements in a (Ga,Mn)As film oriented normal to the magnetic field. We detect the ferromagnetic resonance over a broad frequency range of 2 GHz to 18.5 GHz and determine the spectroscopic $g$-factor and separate the Gilbert from the inhomogeneous contribution to magnetization relaxation. Temperature dependent measurements below the saturation magnetization indicate that the photovoltage signal can serve as a sensitive tool to study the crystal anisotropy. We demonstrate that the combination of spin dynamics with charge transport is a promising tool to study microstructured ferromagnetic semiconductor samples.
\end{abstract}


\maketitle

Ferromagnetic semiconductors \cite{Ohn98Jun06} are currently intensely studied. They offer a unique combination of data processing and data storage in a single material system, which holds great potential for information technology.
Spin excitations such as ferromagnetic resonance (FMR) and standing spin waves (SSW) in (Ga,Mn)As are very attractive both from academic and practical standpoints. They offer access to a broad spectrum of related physics such as the damping mechanisms of magnetization precession \cite{Hei85Gil04,Sin04,Mat06}, many-body effects \cite{Goe03} and the crystal anisotropy \cite{Far98,Liu03,Bih04Lim06,Liu06}.
In (Ga,Mn)As films the FMR and SSW have been studied by absorption experiments \cite{Liu06,Goe03,Liu05a,Liu05b,Liu03,Mat06,Bih04Lim06,Sin04} at fixed frequencies in the X-band (9.47 GHz) and Q-band (34 GHz). Among other insights these nice experiments have provided detailed information on the spectroscopic $g$-factor, the crystal anisotropy and the damping mechanism.
In contrast to absorption experiments, the electric detection of the microwave induced DC photovoltage (PV) enables us to perform broadband measurements of the FMR in (Ga,Mn)As. We cover a frequency range from 2 GHz to 18.5 GHz with a fixed magnetic field orientation.
The DC electric detection of spin excitations is based on the rectification of the induced AC microwave current $\mathbf{j}$ via the Anisotropic Magneto Resistance (AMR). Lately, interest in this technique has increased and it has been applied in Permalloy \cite{Gui07aGui07bMec07}.

Our sample (see Fig. 1) is a Ga$_{1-x}$Mn$_x$As film with $x$ = 2\% and a thickness of $d=$ 50 nm grown on a GaAs substrate.
A hybrid device was fabricated by integrating two Hall bars with a 50 $\Omega$ ground-signal-ground coplanar waveguide (G-S-G CPW), which is connected to a microwave generator and couples the high frequency magnetic field $\mathbf{h}$ and electric field to the Hall bars.
The Hall bars have a width of $w=30$ $\mu$m and a length of $l=0.6$ mm between voltage probes along the $[1\bar{1}0]$ direction.
One Hall bar is centered in each of the two S-G gaps.
The CPW and contacting leads were prepared from 200 nm thick Al.
The geometry of our sample orients the induced high frequency current $\mathbf{j}$ along the $z$-axis.
We estimate the dissipated power at the $^3$He pot of our $^3$He cryostat to be of the order of $P_{dis} \approx 5$ mW from its temperature increase.
Using lock-in measurements at a modulation frequency below 300 Hz we simultaneously measure the PV and using a current of $I=100$ nA the longitudinal and Hall resistances $R_{xx}$ and $R_{xy}$.
The strong T dependence of $R_{xx}$ of (Ga,Mn)As allows us to deduce the true sample temperature under microwave excitation.

\begin{figure} [t]
\begin{center}
\epsfig{file=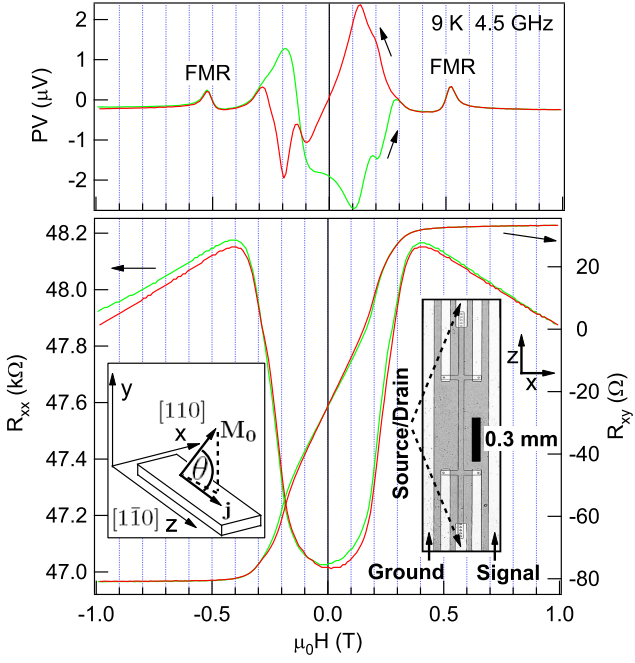,width=8.6 cm} \caption{(color online)
Typical simultaneously measured PV, $R_{xx}$ and $R_{xy}$ for upsweeps (green) and downsweeps (red). The left and right insets show the orientation and a micrograph of the sample. The deviation of the $R_{xx}$ traces on the left is due to a slight drift in temperature of about 1 K.}
\label{Overview}
\end{center}
\end{figure}

Typical simultaneous measurements of the PV, $R_{xx}$ and $R_{xy}$ at 4.5 GHz in magnetic field sweeps are shown in Fig. 1.
The PV exhibits a peak around $\mu_0 |H| = 0.52$ T, which we will later identify as the FMR.
For $\mu_0 |H| \lesssim 0.3 $ T the PV signal shows a strong hysteresis and is much larger.
$R_{xx}$ and $R_{xy}$ also show small hysteretic effects. A discussion of the hysteretic PV will be given later, combining the PV mechanism with the crystal and shape anisotropy.
We assume a slight misalignment of the sample normal ($y$-axis) with the static external field $\mathbf{H}$ ($<0.2^\circ$), since the PV would vanish for parallel orientation of the static magnetization $\mathbf{M_0}$ and the $y$-axis \cite{Gui07aGui07bMec07}.
The Curie temperature $T_C \approx$ 80 K is determined from transport data (not shown).

\begin{figure} [t]
\begin{center}
\epsfig{file=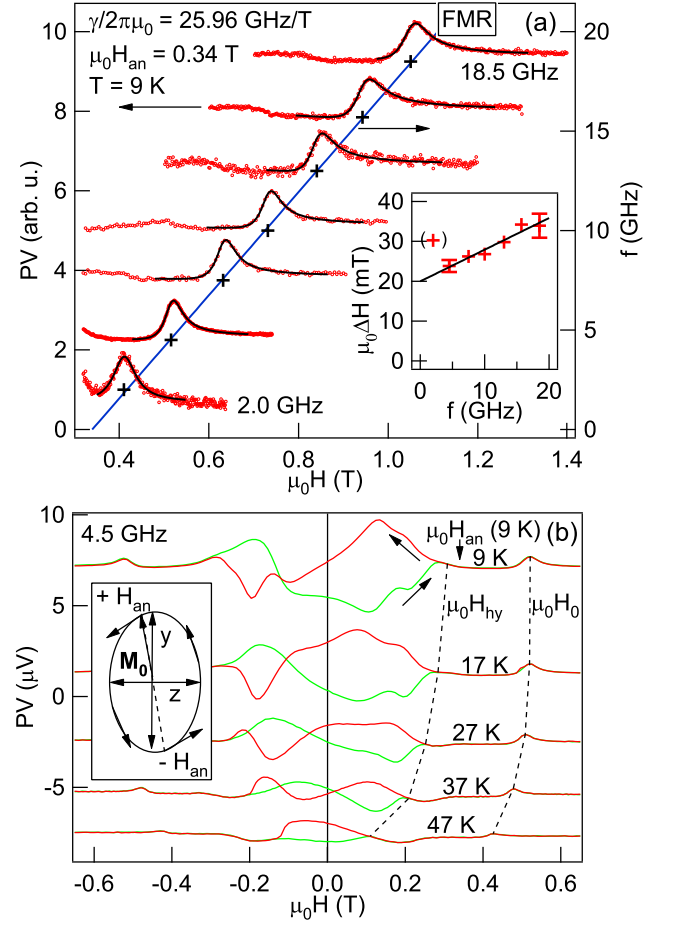,width=8.6 cm} \caption{(color online)
(a) PV signal at different frequencies. The black lines are fits according to Eq. \ref{LandD} and the fitted FMR positions are indicated by crosses. Traces are normalized and offset. The inset shows the FMR linewidth (HWHM) and a linear fit. The value for 2 GHz was excluded from the fit since its lineshape is slightly distorted due to the proximity of $\mu_0 |H_{an}|$.  (b) PV for increasing temperatures for upsweeps (green) and downsweeps (red). The dashed lines are a guide to the eye and traces are offset for clarity. The inset illustrates in a projection to the $z-y$ plane how $\mathbf{M_0}$ may turn during a hysteretic cycle.}
\label{dispersion}
\end{center}
\end{figure}

The lineshape $U(H)$ of the FMR PV can be shown to be a superposition of a symmetric (L) and an antisymmetric (D) Lorentzian  \cite{Gui07aGui07bMec07}
\begin{equation}\label{LandD}
    U(H) = L \frac{\Delta H^2}{(H-H_0)^2 + \Delta H^2} + D \frac{\Delta H (H-H_0)}{(H-H_0)^2 + \Delta H^2}
\end{equation}
where $H_0$ is the resonance field and $\Delta H$ is the half width at half maximum (HWHM) of the symmetric Lorentzian. L and D are determined by the angle $\theta$ between $\mathbf{j}$ and $\mathbf{M_0}$, the orientation and polarization of $\mathbf{h}$ relative to $\mathbf{M_0}$ and the phase of $\mathbf{j}$ relative to $\mathbf{h}$ \cite{Gui07aGui07bMec07}.
Figure 2 (a) shows the PV for different frequencies along with fits according to Eq. \ref{LandD}.
The FMR frequency in the perpendicular configuration can be well described by $\omega = \gamma (|H_0| - M_s + K_{2\bot}/2 \pi M_s)$ \cite{Liu03} where $\gamma$ is the gyromagnetic ratio, $M_s$ the saturation magnetization and $K_{2\bot}$ the uniaxial perpendicular crystal anisotropy parameter. The remaining crystal anisotropy parameters are either not relevant for $\mathbf{M_0}\parallel [001]$ or are of negligible magnitude \cite{Liu03}. In this work we do not measure $M_s$ independently and therefore sum up the shape and crystal anisotropies into a general anisotropy field $H_{an}=M_s - K_{2\bot}/2 \pi M_s$.
We identify the FMR based on the linear dispersion and the slope of our linear fit of $\gamma$ = $d \omega /dH_0$ = $2 \pi \mu_0 $ (25.96 $\pm$ 0.21) GHz/T corresponding to a spectroscopic $g$-factor of $g=1.85$. This value agrees remarkably well with measurements based on the angular dependence of the FMR \cite{Liu05a,Liu05b,Liu06} and thus supports previous evidence for a largely isotropic spectroscopic $g$-factor \cite{Liu05a}.
We deduce $\mu_0 |H_{an}| = 0.34$ T from the zero intercept of the linear fit which agrees well with the value determined from our simultaneously measured transport data \cite{Ohn96Mat98} and also with the value obtained for sample 1 in Ref. 7.
We parenthetically note the weak signal to the low field side of the FMR. This signal is close to the expected position of SSW resonances \cite{Mor01,Goe03}.

The inset of Fig. 2 shows $\Delta H$ extracted from the FMR fits with a linear fit according to $\Delta H = \Delta H_0 + \alpha_G \omega / \gamma$ , with the Gilbert damping factor $\alpha_G$ \cite{Hei85Gil04} and the inhomogeneous contribution $\Delta H_0$.
We obtain $\alpha_G = 0.021 \pm 0.002$, which compares well with previous studies \cite{Sin04,Mat06}.
Our broadband study allows us to separate the intrinsic part of the damping from the inhomogeneous part which is $\mu_0 \Delta H_0 = (20 \pm 1.5)$ mT.
This is a significant advancement, which is made possible by the use of a hybrid device combining broadband, high power excitation with electrical detection and which allows a comparison with theory in future studies \cite{Sin04}. In contrast, previous absorption experiments were limited to a few frequencies.

The PV signal in Fig. 2 (b) shows a decrease of amplitude and $\mu_0 |H_0|$ with increasing temperature, consistent with a decrease of $\mu_0 |H_{an}|$ and with previous absorption measurements \cite{Liu03,Liu06}.
The FMR itself is well suited to study the crystal anisotropy \cite{Liu03,Far98,Bih04Lim06,Liu06}, which is currently of great interest.
Here, we observe a strong temperature dependent hysteretic behavior for $|H|<|H_{hy}|<|H_{an}|$, where $H_{hy}$ is defined as the field below which hysteresis is present.
The relation of the hysteretic behavior to the crystal anisotropy can be qualitatively understood by assuming a slight misalignment of $\mathbf{H}$ and the $y$-axis as depicted in the inset of Fig. 2 (b).
When lowering the field, $\mathbf{M_0}$ will relax from its saturation orientation above $\mu_0 |H_{an}|$ towards the in-plane easy crystal axes and be forced out of plane again, as $\mathbf{H}$ is reversed (downseep).
In the following upsweep, $\mathbf{M_0}$ is likely to relax back to the in plane easy crystal axes in opposite orientation compared to the downsweep \cite{Liu05b}.
The resulting hysteresis of the PV is strikingly large compared to the transport data since the PV can change its sign and is in general very sensitive to $\mathbf{M_0}$ and $\theta$ \cite{Gui07aGui07bMec07}.
A detailed quantitative understanding of the hysteretic PV signal may reveal the path of $\mathbf{M_0}$ and therefore the crystal anisotropies.
This understanding can be based on the theory of the microwave PV \cite{Gui07aGui07bMec07}, which was recently used to detect $\mathbf{h}$ for a known $\mathbf{M_0}$ and $\mathbf{j}$ \cite{Bai08}.
The PV shows points where up- and downsweeps cross, which may hold further information on $\mathbf{M_0}$, since they imply either an identical $\mathbf{M_0}$ or a symmetric orientation of $\mathbf{M_0}$ with respect to the PV amplitude.
We note that the resonant PV for $|H|<|H_{an}|$ is most pronounced in an interval of about 1 GHz around 4.5 GHz. This frequency is determined by the shape and crystal anisotropies, which provide a restoring force for the FMR even at $\mu_0 H = 0$ T.

The demonstrated electrical detection of spin excitations using the PV technique is a promising tool to study microstructured samples with small absorption, and may thus prove valuable in areas such as time domain studies of magnetization dynamics, domain wall motion, spin wave propagation and magnetic domain wall logic. The PV can also be used as a frequency filter and a microscopic voltage source.

In summary, we have detected the FMR in (Ga,Mn)As in a broad frequency range from 2 GHz to 18.5 GHz by means of the PV technique. From this data we extracted the spectroscopic $g$-factor, the intrinsic Gilbert damping factor $\alpha_G$ and the inhomogeneous damping contribution. We have observed a striking hysteresis of the PV as a function of temperature and discussed the possibility to study the crystal anisotropy in (Ga,Mn)As by analyzing this hysteresis.

We thank G. Roy and G. Mollard for technical assistance.
This work has been funded by NSERC, CFI, and URGP grants awarded to C.-M. Hu. The hybrid device was fabricated in the Nano-Systems Fabrication Lab at the University of Manitoba.

\end{document}